\documentclass[showkeys,showpacs,superscriptaddress,nofootinbib]{revtex4-2}
\usepackage{amsmath}
\usepackage{amssymb}
\usepackage{dutchcal}
\usepackage{graphicx}
\usepackage[normalem]{ulem}
\usepackage{xcolor}

\begin{document}

\title{Wormholes in Einstein-Dirac-Maxwell theory \\ with identical spacetime asymptotics 
}
\author{
Vladimir Dzhunushaliev
}
\email{v.dzhunushaliev@gmail.com}
\affiliation{
Department of Theoretical and Nuclear Physics,  Al-Farabi Kazakh National University, Almaty 050040, Kazakhstan
}
\affiliation{
Institute for Experimental and Theoretical Physics, Al-Farabi Kazakh National University, Almaty 050040, Kazakhstan
}
\affiliation{Academician J.~Jeenbaev Institute of Physics of the NAS of the Kyrgyz Republic, 265 a, Chui Street, Bishkek 720071, Kyrgyzstan}

\author{Vladimir Folomeev}
\email{vfolomeev@mail.ru}
\affiliation{Academician J.~Jeenbaev Institute of Physics of the NAS of the Kyrgyz Republic, 265 a, Chui Street, Bishkek 720071, Kyrgyzstan}

\author{Nurzada Beissen}
\affiliation{
Department of Theoretical and Nuclear Physics,  Al-Farabi Kazakh National University, Almaty 050040, Kazakhstan
}
\affiliation{
Institute for Experimental and Theoretical Physics, Al-Farabi Kazakh National University, Almaty 050040, Kazakhstan
}

\author{Adilet Nurmukhamedov}
\affiliation{
Department of Theoretical and Nuclear Physics,  Al-Farabi Kazakh National University, Almaty 050040, Kazakhstan
}

\begin{abstract}
Within general relativity, we study spherically symmetric configurations with wormhole topology consisting of spinor fields and a Maxwell electric field.
For such a system, we construct complete families of regular asymmetric solutions describing wormholes connecting two identical Minkowski spacetimes.
The physical properties of such systems are completely determined by the values of three input quantities: the throat parameter, the spinor frequency, 
and the coupling constant. Depending on the specific values of these parameters, the configurations may 
have essentially different characteristics, including negative ADM masses.
\end{abstract}


\date{\today}

\maketitle 

\section{Introduction}

Wormholes are hypothetical objects possessing a nontrivial spacetime topology and connecting either two distant points in one universe or
even different universes~\cite{Visser}. In order to construct solutions describing such objects in general relativity, 
it is usually necessary to have some form of exotic matter violating the null/weak energy conditions.
As such exotic matter, it is possible to employ, for example, 
one of the forms of dark energy, which, according to the contemporary viewpoint, ensures the 
acceleration of the present Universe~\cite{AmenTsu2010}. The observational data indicate~\cite{Ade:2015xua}
that dark energy violating the null/weak energy conditions may also exist in the Universe. 

In the simplest case, the exotic matter can be modeled by
the so-called ghost (or phantom) scalar fields, which may either be massless~\cite{Bronnikov:1973fh,Ellis:1973yv,Ellis:1979bh}
or have a potential energy~\cite{Kodama:1978dw,Kodama:1978zg}. However, while remaining within general relativity, 
one can nevertheless try to avoid the use of any phantom matter. An interesting attempt to find such solutions
is a consideration of static, {\it symmetric} with respect to the wormhole throat solutions 
obtained within Einstein-Dirac-Maxwell  theory~\cite{Blazquez-Salcedo:2020czn}. However, those solutions have certain features
(for instance, the introduction of a thin shell is needed) that were criticized in the papers~\cite{Bolokhov:2021fil,Danielson:2021aor,Konoplya:2021hsm}.
As a possibility to avoid the weaknesses of the model of Ref.~\cite{Blazquez-Salcedo:2020czn}, 
the pioneering work~\cite{Konoplya:2021hsm}, also within Einstein-Dirac-Maxwell  theory, 
suggests static {\it asymmetric} wormholes supported by smooth metric and matter (classical spinor and electric) fields with no any thin shells.
The corresponding solutions are sought using the shooting method with boundary conditions imposed at the center of the configuration 
(in particular, it is assumed that the spatial derivative of the metric function $g_{tt}$ is equal to zero at the center). 
This enables one to obtain solutions describing a wormhole connecting two {\it nonidentical} Minkowski spacetimes. 
 In turn, the authors of the paper~\cite{Wang:2022aze} considered the case where the spatial derivative of  $g_{tt}\neq 0$ at the center;
 then, solving the problem with boundary conditions imposed both at the center and at infinity, they found  
 another family of solutions which, however, also describe  two {\it nonidentical } asymptotically flat spacetimes.
 
In the present paper we work within the framework of the model~\cite{Konoplya:2021hsm} and study one more possibility when
 {\it all} boundary conditions are imposed at plus/minus infinity. The solution of such a two-point boundary value problem enables us
 to get regular solutions that already describe  two {\it  identical } Minkowski spacetimes.
 As a result, we obtain a new family of solutions parameterized by three free system parameters: 
 the throat parameter (which determines the geometry of a spacetime in the vicinity of the wormhole's throat),
the spinor frequency (i.e., the frequency of oscillations of the spinor field),
and the coupling constant (which characterizes the interaction between the spinor field and the electric field).

The paper is organized as follows. In Sec.~\ref{prob_statem}, we give the general-relativistic equations for the configurations under consideration.
These equations are solved numerically in Sec.~\ref{num_sol} for the neutral spinor field
(Sec.~\ref{sec_e_0}) and for the charged spinor field (Sec.~\ref{sec_e_neq_0}). 
Then we find some approximate analytical solutions for the systems under consideration in Sec.~\ref{sec_approx_sol} and discuss the question of stability in Sec.~\ref{sec_M_R_stab}. 
Finally, in Sec.~\ref{concl}, we summarize the results obtained.

\section{Formulation of the problem and general equations}
\label{prob_statem}

The total action for the system can be written in the form [we use the metric signature $(+,-,-,-)$ and natural units $c=\hbar=1$]
\begin{equation}
\label{action_gen}
	S_{\text{tot}} = - \frac{1}{16\pi G}\int d^4 x
		\sqrt{-\cal{g}} R +S_{\text{sp}} +S_{\text{EM}},
\end{equation}
where $G$ is the Newtonian gravitational constant, $R$ is the scalar curvature, and $\cal{g}$ is the determinant of the metric;
   $S_{\text{sp}}$ and $S_{\text{EM}}$ denote the actions of spinor, $\psi$,
and electromagnetic, $A_\mu$, fields, respectively.
The action for the electromagnetic field can be found from the Lagrangian 
$$
L_{\text{EM}}=-\frac{1}{4}F_{\mu\nu}F^{\mu\nu},
$$
where the electromagnetic field tensor is $F_{\mu\nu}=\partial_\mu A_\nu-\partial_\nu A_\mu$ with $\mu, \nu = 0, 1, 2, 3$ being spacetime indices.

In turn, the action  $S_{\text{sp}}$ for the spinor field $\psi$ appearing in Eq.~\eqref{action_gen} can be obtained from the Lagrangian 
$$
	L_{\text{sp}} =	\frac{\imath}{2} \left(
			\bar \psi \gamma^\mu \psi_{; \mu} -
			\bar \psi_{; \mu} \gamma^\mu \psi
		\right) - \mu \bar \psi \psi ,
$$
where $\mu$ is the mass of the spinor field and the semicolon denotes the covariant derivative defined as
$
\psi_{; \mu} =  [\partial_{ \mu} +1/8\, \omega_{a b \mu}\left( \gamma^a  \gamma^b- \gamma^b  \gamma^a\right) - \imath e A_\mu]\psi 
$. 
Here $\gamma^a$ are the Dirac matrices in the standard representation in flat space 
 [see, e.g.,  Ref.~\cite{Lawrie2002}, Eq.~(7.27)]. In turn, the Dirac matrices in curved space, $\gamma^\mu = e_a^{\phantom{a} \mu} \gamma^a$, are derived  using the tetrad $ e_a^{\phantom{a} \mu}$, and $\omega_{a b \mu}$ is the spin connection
[for its definition, see Ref.~\cite{Lawrie2002}, Eq.~(7.135)].
The gauge coupling constant $e$ describes the minimal interaction between electromagnetic
 and spinor fields.

Then, by varying the action \eqref{action_gen} with respect to the metric, the spinor field, and the vector potential $A_\mu$, we derive the Einstein, Dirac, and Maxwell field equations in curved spacetime:
\begin{eqnarray}
G_{\mu}^\nu	\equiv R_{\mu}^\nu - \frac{1}{2} \delta_{\mu }^\nu R &=&
	8 \pi  G  \,T_{\mu }^\nu,
\label{feqs_10} \\
	\imath \gamma^\mu \psi_{;\mu} - \mu  \psi &=& 0 ,
\label{feqs_20}\\
	\imath \bar\psi_{;\mu} \gamma^\mu + \mu  \bar\psi &=&0 ,
\label{feqs_30}\\
\frac{1}{\sqrt{-\cal{g}}} \frac {\partial}{\partial x^\mu}
    \left(\sqrt{-\cal{g}}F^{\mu \nu}\right) &=& -e j^{\nu} ,
\label{feqs_40}
\end{eqnarray}
where $ j^{\nu}= \bar{\psi} \gamma^\nu \psi$ is the current of the spinor field.
The equation~\eqref{feqs_10} involves the energy-momentum tensor $T_{\mu}^\nu$, which can be written in a symmetric form as
\begin{equation}
\label{EM_1}
	T_{\mu}^\nu = \frac{\imath }{4}g^{\nu\rho}
	\left[
		\bar\psi \gamma_{\mu} \psi_{;\rho} 
		+ \bar\psi\gamma_\rho\psi_{;\mu} - \bar\psi_{;\mu}\gamma_{\rho }\psi 
		- \bar\psi_{;\rho}\gamma_\mu\psi
	\right] - F^{\nu\rho} F_{\mu\rho}
    + \frac{1}{4} \delta_\mu^\nu F_{\alpha\beta} F^{\alpha\beta}.
\end{equation}

Since we consider here only spherically symmetric configurations,  one can choose the spacetime metric in the form
\begin{equation}
	ds^2 = e^{A(r)} dt^2 - B(r) e^{-A(r)}\left[dr^2 + \left(r^2+r_0^2\right) \left(d \theta^2 + \sin^2 \theta d \varphi^2\right)\right],
\label{metric}
\end{equation}
where $e^{A(r)}=1-2 G m(r)/r$ with the function $m(r)$ corresponding to the current mass of the configuration
enclosed by a sphere with circumferential radius $r$, and the parameter $r_0$ characterises the throat.

For a description of the spinor field, we take the following stationary {\it Ansatz} 
compatible with the spherically symmetric line element \eqref{metric} (see, e.g., Refs.~\cite{Soler:1970xp,Li:1982gf,Li:1985gf,Herdeiro:2017fhv}):
\begin{equation}
	\psi =  
	\frac{1}{2} e^{-\imath t \Omega }
	\begin{pmatrix}
	 -\imath e^{\frac{1}{2} \imath (\theta -\varphi )} u(r) 	& -e^{\frac{1}{2} \imath (\theta +\varphi )} u(r) \\
	 \imath e^{-\frac{1}{2} \imath (\theta +\varphi )} u(r)  & -e^{-\frac{1}{2} \imath (\theta -\varphi )} u(r) \\
	 -e^{-\frac{1}{2} \imath (\theta +\varphi )} v(r) 	& -\imath e^{-\frac{1}{2} \imath (\theta -\varphi )} v(r) \\
	 e^{\frac{1}{2} \imath (\theta -\varphi )} v(r) 			& -\imath e^{\frac{1}{2} \imath (\theta +\varphi )} v(r) \\
	\end{pmatrix} ,
\label{spinor} 
\end{equation}
where $\Omega$ is the spinor frequency and $u(r)$ and $v(r)$ are two real functions. 
This  {\it Ansatz} ensures that the spacetime remains static. Each row of this  {\it Ansatz} describes a  spin-$\frac{1}{2}$ fermion, 
and these two fermions have the same masses~$\mu$ and opposite spins. Even though the energy-momentum tensors of these fermions are not spherically symmetric, their sum ensures a spherically symmetric energy-momentum tensor.

The gauge field is parameterized by an electric potential
\begin{equation}
A_\mu=\{\phi(r),0,0,0\}.
\label{EM_ans}
\end{equation}

Then, substituting the {\it Ans\"{a}tze} \eqref{spinor} and \eqref{EM_ans}  and the metric  \eqref{metric} in the field equations \eqref{feqs_10}, \eqref{feqs_20}, and \eqref{feqs_40}, one can obtain the following set of equations:
\begin{align}
& A^{\prime\prime}+\frac{1}{2}\left(\frac{4 x}{x^2+x_0^2}-A^\prime\right)A^\prime+\frac{1}{2}\left(\frac{4 x}{x^2+x_0^2}+A^\prime+\frac{B^\prime}{B}\right)\frac{B^\prime}{B}
-\frac{2 x_0^2}{\left(x^2+x_0^2\right)^2}+\frac{8 e^{-A/2}\sqrt{B}\bar u\bar v}{\sqrt{x^2+x_0^2}}\nonumber \\
&+2 e^{-3 A/2} B \left[\left(3 e^{A/2}-4 \bar{U}\right)\bar{u}^2-\left(3 e^{A/2}+4 \bar{U}\right)\bar{v}^2\right]=0,
\label{Eq_A}\\
& B^{\prime\prime}+\left(\frac{3 x}{x^2+x_0^2}-\frac{1}{2}\frac{B^\prime}{B}\right)B^\prime+4 e^{-A/2}B^{3/2}\frac{\bar u \bar v}{\sqrt{x^2+x_0^2}}+
4 e^{-3 A/2}B^2\left[\left( e^{A/2}- \bar{U}\right)\bar{u}^2-\left( e^{A/2}+ \bar{U}\right)\bar{v}^2\right]=0,
\label{Eq_B}\\
& \bar u^\prime+\left(\frac{x}{x^2+x_0^2}-\frac{1}{\sqrt{x^2+x_0^2}}-\frac{1}{4}A^\prime+\frac{1}{2}\frac{B^\prime}{B}\right)\bar u+
e^{-A}\sqrt{B}\left( e^{A/2}+ \bar{U}\right)\bar{v}=0,
\label{Eq_u}\\
& \bar v^\prime+\left(\frac{x}{x^2+x_0^2}+\frac{1}{\sqrt{x^2+x_0^2}}-\frac{1}{4}A^\prime+\frac{1}{2}\frac{B^\prime}{B}\right)\bar v+
e^{-A}\sqrt{B}\left( e^{A/2}- \bar{U}\right)\bar{u}=0,
\label{Eq_v}\\
& \bar{U}^{\prime\prime}+\left(\frac{2 x}{x^2+x_0^2}-A^\prime+\frac{1}{2}\frac{B^\prime}{B}\right)\bar{U}^{\prime}- \bar{e}^2 e^{-A/2} B\left(\bar u^2+\bar v^2\right)=0 ,
\label{Eq_Maxw}
\end{align}
where the prime denotes differentiation with respect to the radial coordinate and we have introduced an auxiliary function
  $\bar{U}=\bar{\Omega}+\bar{e} \bar{\phi}$.
The  equations~\eqref{Eq_A} and ~\eqref{Eq_B} are obtained using the  combinations $\left[\left(_t^t\right)-\left(_r^r\right)\right]$ and $\left[\left(_t^t\right)+\left(_\theta^\theta\right)\right]$
of the Einstein equations~\eqref{feqs_10}.
Notice that these equations are invariant with respect to multiplying the spinor functions by -1 and with respect to simultaneous changes $\bar{\phi}\to -\bar{\phi}$ and $\bar{e}\to -\bar{e}$.
The above equations are written in terms of the following dimensionless variables and parameters:
\begin{equation}
\label{dmls_var}
	x =  \mu r, \quad
	\bar \Omega = \frac{\Omega}{\mu}, \quad
	\left(\bar u, \bar v\right) = \sqrt{\frac{4\pi G}{\mu}}\left( u, v\right),\quad
	  \bar \phi = \sqrt{4\pi G}\phi,\quad \bar{e}=\frac{e}{\mu \sqrt{4\pi G}}
		 .  
\end{equation}

Notice here that the first integral of the Maxwell equation \eqref{Eq_Maxw} is
\begin{equation}
	\bar{U}^\prime =  \frac{ \bar{e} \,e^A}{\sqrt{B}\left( x^2 + x_0^2\right) } \left[\bar{Q}
	+ \bar{e} \int_{x_{\text{max}}}^{x} 
e^{-3 A/2}B^{3/2}\left( x^{2} + x_0^2\right) \left(\bar u^2 + \bar v^2\right) d x \right] ,
\label{Maxw_sol}
\end{equation}
where $\bar Q \equiv  \sqrt{G/4\pi}\mu \,Q$ is an integration constant which represents  the charge of a sourceless electric field,
and this constant can be found from Eq.~\eqref{Maxw_sol} in the form
$$
	 \bar{Q} = \frac{1}{\bar{e}}\left. \left[ 
		\left(x^2 + x_0^2\right) e^{-A} \sqrt{B}\,\bar{U}^\prime
	\right] \right|_{x = x_{\text{max}}} .
$$
The equation \eqref{Maxw_sol} implies that the lines of force of the electric field originate as $x \to - \infty$ (or $x \to + \infty$) 
and end as $x \to + \infty$ (or $x \to - \infty$), respectively;
such a situation is possible because we are dealing with a wormhole geometry.
In turn, the integral in Eq.~\eqref{Maxw_sol} is associated with the Noether charge of the spinor fileld (see Eq.~\eqref{part_num}
where the definition of the lower limit of integration $x_{\text{max}}$ is also given).

The set of equations \eqref{Eq_A}-\eqref{Eq_Maxw} is supplemented by the constraint equation [the $\left(^r_r\right)$-component of the Einstein equations~\eqref{feqs_10}]
\begin{align}
\label{eq_constr}
&\frac{e^{A}}{4 B}\left[A^{\prime 2}-\left(\frac{B^{\prime }}{B}+\frac{4 x}{x^2+x_0^2}\right)\frac{B^\prime}{B}+\frac{4x_0^2}{\left(x^2+x_0^2\right)^2}\right]	-\frac{\bar{U}^{\prime 2}}{\bar{e}^2 B}-
\frac{4 e^{A/2}\bar u \bar v}{\sqrt{B\left(x^2+x_0^2\right)}}\nonumber \\
&+2 e^{-A/2}\left[-\left(e^{A/2}-\bar{U}\right)\bar{u}^2+\left(e^{A/2}+\bar{U}\right)\bar{v}^2\right]=0 .
\end{align}
Below we employ this equation to determine the boundary conditions for the electric field in the case with $\bar{e}\neq 0$.

\section{Numerical solutions}
\label{num_sol}

In this section, we numerically solve the equations~\eqref{Eq_A}-\eqref{Eq_Maxw}  and discuss the physical properties of the configurations obtained.

\subsection{Asymptotic behavior and boundary conditions}

We seek regular finite-energy solutions of five ordinary differential equations~\eqref{Eq_A}-\eqref{Eq_Maxw}.
Even before numerical solution of these equations, it is possible to estimate an asymptotic behavior of the solutions, bearing in mind that
the spinor fields $\bar u$ and $\bar v$  are functions, which diminish exponentially as $x\to \pm \infty$. 
It is seen from the form of Eqs.~\eqref{Eq_u} and \eqref{Eq_v} that, due to the presence of the term $\left(x^2+x_0^2\right)^{-1/2}$, they are not $Z_2$-symmetric. 
(Notice that such an asymmetry is inherent not only to the wormhole solutions considered here but also to the domain wall solutions supported by a spinor field~\cite{Dzhunushaliev:2023sdq}.)
This implies that we should seek solutions that are asymmetric with respect to the origin of coordinates $x=0$.
Then the corresponding asymptotic behavior of the spinor fields has the form
\begin{equation}
\bar{u}\approx \bar{u}_{\pm\infty}\frac{e^{\mp\sqrt{1-\bar{U}_{\pm\infty}^2}x}}{x}+\cdots, 
\quad \bar{v}\approx \bar{v}_{\pm\infty}\frac{e^{\mp\sqrt{1-\bar{U}_{\pm\infty}^2}x}}{x}+\cdots,
\label{u_v_asympt}
\end{equation}
where $\bar{u}_{\pm\infty}$ and $\bar{v}_{\pm\infty}$ are integration constants as $x\to \pm \infty$, respectively,
and $\bar{U}_{\pm\infty}$ are asymptotic values of the function $\bar{U}$ as $x\to \pm \infty$. 
For the uncoupled fermions ($\bar e=0$), $\bar{U}_{\pm\infty}$ correspond to the spinor frequency $\bar{\Omega}$.

In turn, for the metric functions $A$ and $B$, one can find from Eqs.~\eqref{Eq_A} and~\eqref{Eq_B} as  $x\to \pm\infty$
\begin{equation}
 A\approx \mp\frac{2 \bar M_\pm}{x}+\cdots, \quad B \approx 1+\frac{\alpha}{x^2}+\cdots .
\label{A_asympt}
\end{equation}
Here the dimensionless $\bar M_\pm\equiv \mu M_\pm/M_p^2$ with $M_p$ being the Planck mass, and 
$\bar M_+$ corresponds to the ADM (Arnowitt-Deser-Misner) mass of the configurations under consideration
as measured by a distant observer when $x\to +\infty$ and $\bar{M}_-$ is the mass as measured when $x\to -\infty$;
$\alpha$ is a constant.

The asymptotic behavior of the electric field follows from the Maxwell equation~\eqref{Eq_Maxw} in the form
$$
\bar{\phi}\approx \bar{\phi}_{\pm \infty}-\frac{\bar{Q}_\pm}{x} +\cdots ,
$$
where $\bar{\phi}_{\pm \infty}$ are two integration constants corresponding to the values of the field as $x\to \pm \infty$, respectively,
and $\bar Q_\pm\equiv  \sqrt{G/4\pi}\mu \,Q_\pm$ represents the  charge of the systems located to the left ($\bar Q_-$) and to the right ($\bar Q_+$) of the center.
Since the integration constants $\bar{\phi}_{\pm \infty}$ are arbitrary, for calculations given below, we choose the constant  $\bar{\phi}_{-\infty}=0$;
this in turn implies that $\bar{\phi}_{+\infty}=\left(\bar{U}_{+\infty}-\bar{U}_{-\infty}\right)/\bar{e}$. In this case $\bar{U}_{-\infty}$ corresponds to the spinor frequency $\bar{\Omega}$.

Then, on account of the asymptotic behavior given above, the corresponding boundary conditions can be taken in the form
\begin{equation}
\label{BCs}
A(x\to \pm \infty)=0, \quad B(x\to \pm \infty)=1, \quad \bar u(x\to \pm \infty)=0, \quad \bar v(x\to \pm \infty)=0, \quad \bar \phi(x\to \pm \infty)=\bar{\phi}_{\pm \infty}.
\end{equation}
These boundary conditions imply that we are dealing with two identical Minkowski spacetimes (unlike the solutions considered in Refs.~\cite{Konoplya:2021hsm,Wang:2022aze}
where at least one of the asymptotic ends is not Minkowski spacetime). 

\subsection{Numerical method}

We solve the set of mixed order differential equations~\eqref{Eq_A}-\eqref{Eq_Maxw} 
with the boundary conditions~\eqref{BCs} and verify that the constraint equation \eqref{eq_constr}
 is satisfied (this is achieved by an appropriate choice of the asymptotic value $\bar{U}_{+\infty}$).

In order to map the infinite range of the radial variable $x$ 
to the finite interval, we introduce the compactified coordinate~$\bar x$ as follows,
\begin{equation}
     x = c_k\frac{\bar x}{\left(1-\bar x^2\right)^2} \, ,
\label{comp_coord}
\end{equation}
which maps the infinite region $(-\infty;\infty)$ onto the finite interval $[-1; 1]$. Here $c_k$ is a constant which is used to adjust the contraction of
the grid. In our calculations, we typically take $c_k\in [0.1,1.3]$.

Technically, Eqs.~\eqref{Eq_A}-\eqref{Eq_Maxw}  are discretized on a grid
consisting of about 1000 grid points, but in some cases even 3000 or more grid points have been used.
The resulting system of nonlinear algebraic equations 
is then solved by using a modified Newton method.
The underlying linear system is solved 
with the Intel MKL PARDISO sparse direct solver~\cite{pardiso} 
and the CESDSOL library\footnote{Complex Equations-Simple Domain 
partial differential equations SOLver, a C++ package developed by I.~Perapechka,
see Refs.~\cite{Herdeiro:2019mbz,Herdeiro:2021jgc}.}.
The package provides an iterative procedure to obtain an exact solution starting from some initial guess configuration. 
As such a configuration, we take a system found in Ref.~\cite{Konoplya:2021hsm}.

\subsection{Mass and the circumferential radius}

We consider configurations possessing a nontrivial spacetime topology that are asymptotically flat and
asymmetric with respect to the center $x=0$. The important point here is the behavior 
of the circumferential radius $\bar R(x)$ defined as
\begin{equation}
\bar R^2\equiv g_{\theta\theta}=B e^{-A}\left(x^2+x_0^2\right) .
\label{circ_radius}
\end{equation} 
Asymptotic flatness implies that $\bar R(x) \to |x|$ for large $|x|$.
Due to the  asymmetry of the configurations, 
the center of the systems located at $x=0$ should not in general be an extremum
of $\bar R(x)$. Depending on the concrete values of the system parameters, the  extremum
of $\bar R(x)$ can be resided both to the left and to the right of the point $x=0$. 
If $\bar R(x)$ has only one global minimum at  some point $x=x_{\text{extr}}$, then $x_{\text{extr}}$ is the throat
of the wormhole $\bar R_{\text{th}}=\text{min}\{\bar R(x)\}$ (a single-throat system).
If, on the other hand, $\bar R(x)$ has a local maximum at  $x=x_{\text{extr}}$,
then this point is an equator $\bar R_{\text{eq}}=\text{max}\{\bar R(x)\}$.
This then implies that there are (at least) two minima of $\bar R(x)$ (on account of the asymmetry of the system,
one of them is global and another one~-- local), located in general asymmetrically to the left and to the right of the maximum. 
In the case of two such minima, the wormhole has a double throat
 (see, e.g., Refs.~\cite{Charalampidis:2013ixa,Hauser:2013jea,Dzhunushaliev:2014mza,Dzhunushaliev:2025fbf,Dzhunushaliev:2025qhw}).

Let us now turn to the total mass of the systems under consideration. Since they are asymmetric with respect to the center $x=0$,
a distant observer placed at $x\to \pm \infty$ measures different magnitudes of the masses $\bar{M}_+$ and $\bar{M}_-$ at plus/minus infinity, see Eq.~\eqref{A_asympt}.
They correspond  to a dimensionless ADM mass of the system $\bar M$ given by
\begin{equation}
\label{expres_mass}
\bar M_\pm\equiv \mu M_\pm/M_p^2=\pm\frac{1}{2}\lim_{x\to\pm\infty}x^2\partial_x A = \pm\frac{c_k}{8}\lim_{\bar x\to \pm 1}\frac{\partial_{\bar x} A }{1-\bar x^2},
\end{equation}
where  the last expression in the above equation represents the mass in terms of the
compactified coordinate $\bar x$ from Eq.~\eqref{comp_coord}.

Alternatively, the ADM mass of the configuration may also be found from the $(^t_t)$-component
of the energy-momentum tensor~\eqref{EM_1}. In the case of spherical symmetry, the Misner-Sharp~\cite{Misner:1964je} mass $m(r)$ associated with the volume enclosed
by a sphere with circumferential radius $R_{\text{extr}}$ and another sphere with
circumferential radius $R > R_{\text{extr}}$ can be defined as follows:
\begin{equation}
m(r)=\frac{1}{2G} R_{\text{extr}}+4\pi\int_{R_{\text{extr}}}^r T_t^t R^2 dR .
\label{m_current}
\end{equation}
In this expression, the  circumferential radius $R_{\text{extr}}$ corresponds either to 
the radius of the wormhole throat  $R_{\text{th}}$ (for single-throat systems) or to the radius of the equator $R_{\text{eq}}$
(for double-throat systems). Taking the boundary to (spacelike) infinity, the Misner-Sharp mass leads to the ADM mass.
In terms of the dimensionless variables~\eqref{dmls_var} the formula \eqref{m_current} can be represented in the form
\begin{equation}
\bar m(x)\equiv  \mu m(r)/M_p^2=\frac{\bar{R}_{\text{extr}}}{2}+\int_{x_{\text{extr}}}^{x}\bar T_t^t \bar R^2 \frac{d\bar R}{d x^\prime} d x^\prime ,
\label{m_current_dmls}
\end{equation}
where $x_{\text{extr}}$ is the point on the $x$-axis where a throat or an equator are located.
The expression~\eqref{m_current_dmls} can be used to monitor the accuracy of the numerical calculations.
In turn, using the metric~\eqref{metric} and the   {\it Ans\"{a}tze} \eqref{spinor} and \eqref{EM_ans}, 
the  $(^t_t)$-component of the energy-momentum tensor~\eqref{EM_1} appearing here can be found in the form
\begin{equation}
\bar T_t^t=\frac{1}{2}\frac{\bar{\phi}^{\prime 2}}{B}+e^{-A/2}\left(\bar{u}^2+\bar{v}^2\right)\bar{U} .
\label{Ttt_dmls}
\end{equation}

\subsection{Results of numerical calculations}

The boundary conditions in the form~\eqref{BCs} imply the solution of a two-point boundary value problem for the set of equations~\eqref{Eq_A}-\eqref{Eq_Maxw},
when the constraint equation~\eqref{eq_constr} is taken into account.
This set is solved by assigning different values of the free system parameters  $x_0$,  $\bar\Omega$, and $\bar{e}$ for which one can find regular solutions.

\subsubsection{The case of $\bar{e}=0$}
\label{sec_e_0}

\begin{figure}[t]
    \begin{center}
        \includegraphics[width=.49\linewidth]{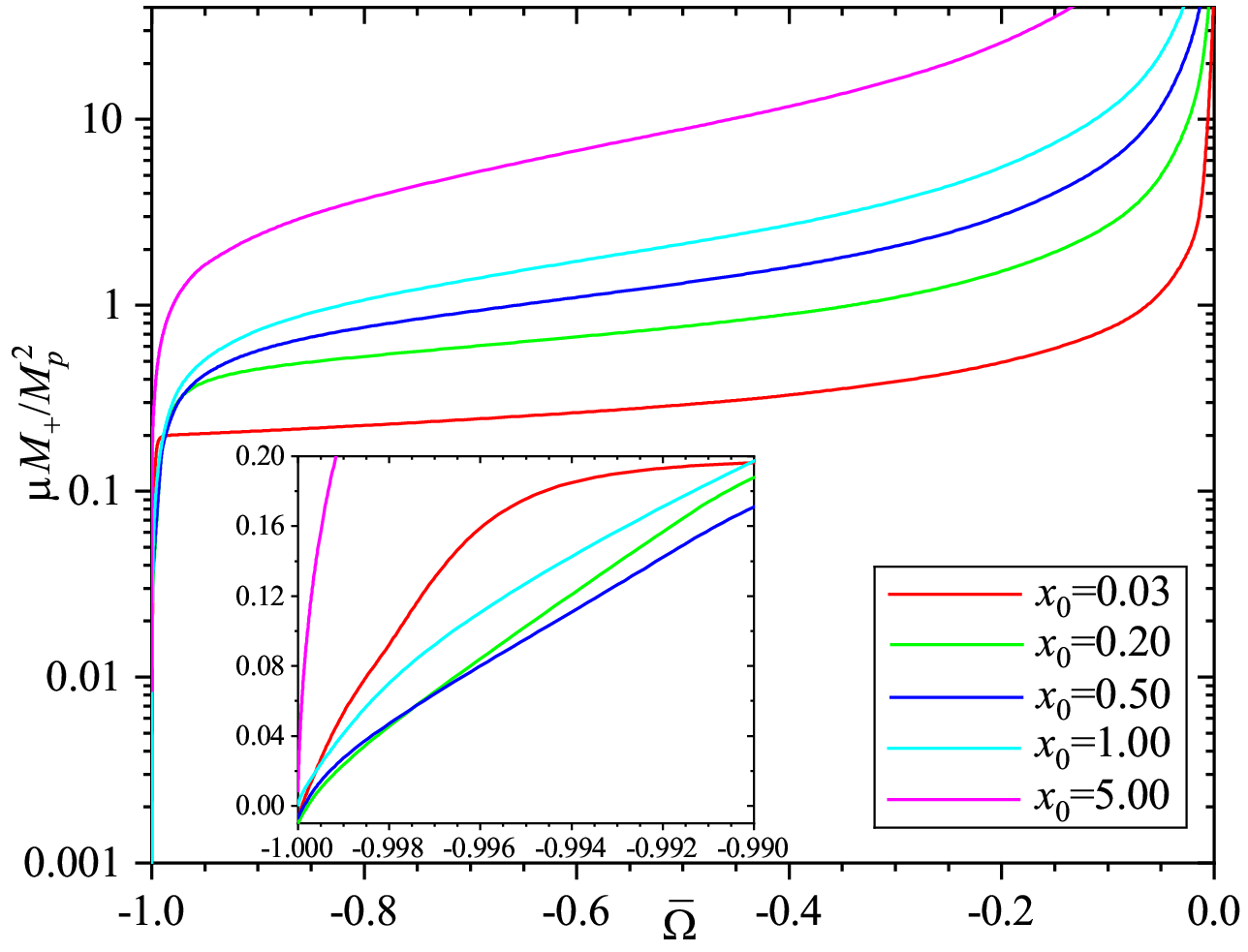}
        \includegraphics[width=.49\linewidth]{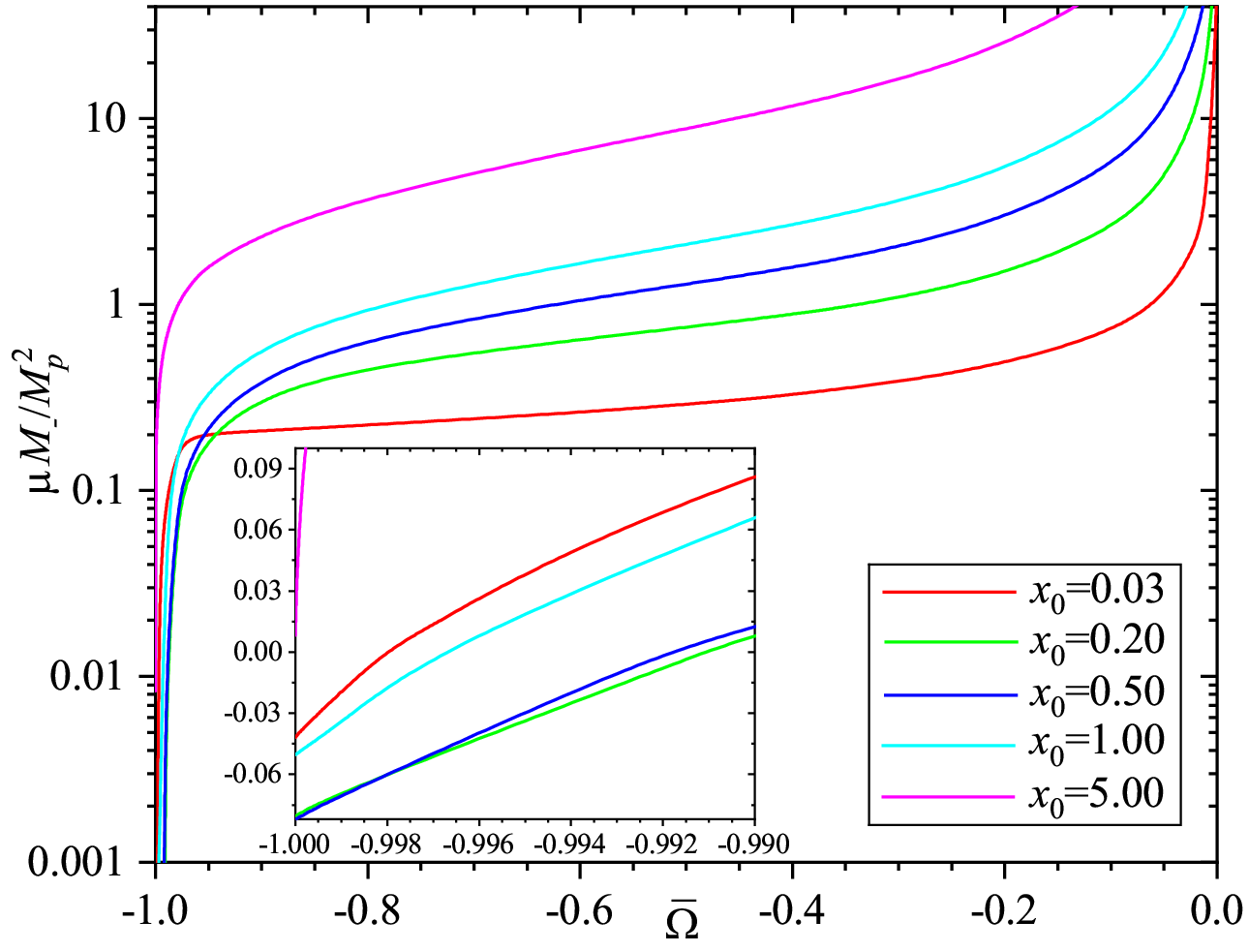}
    \end{center}
    \vspace{-.5cm}
    \caption{The dimensionless total masses $\bar M_\pm$ as functions of
the parameter $\bar\Omega$ for an uncharged spinor field ($\bar{e}=0$) and different values of $x_0$. 
As   $\bar\Omega\to 0$, the masses  diverge as $\bar M_\pm\sim \bar\Omega^{-1}$.
The insets show the behavior of the curves in the region $\bar{\Omega}\to -1$.}
    \label{fig_Mass_Omega_e_0}
\end{figure}

\begin{figure}[!]
    \begin{center}
        \includegraphics[width=.5\linewidth]{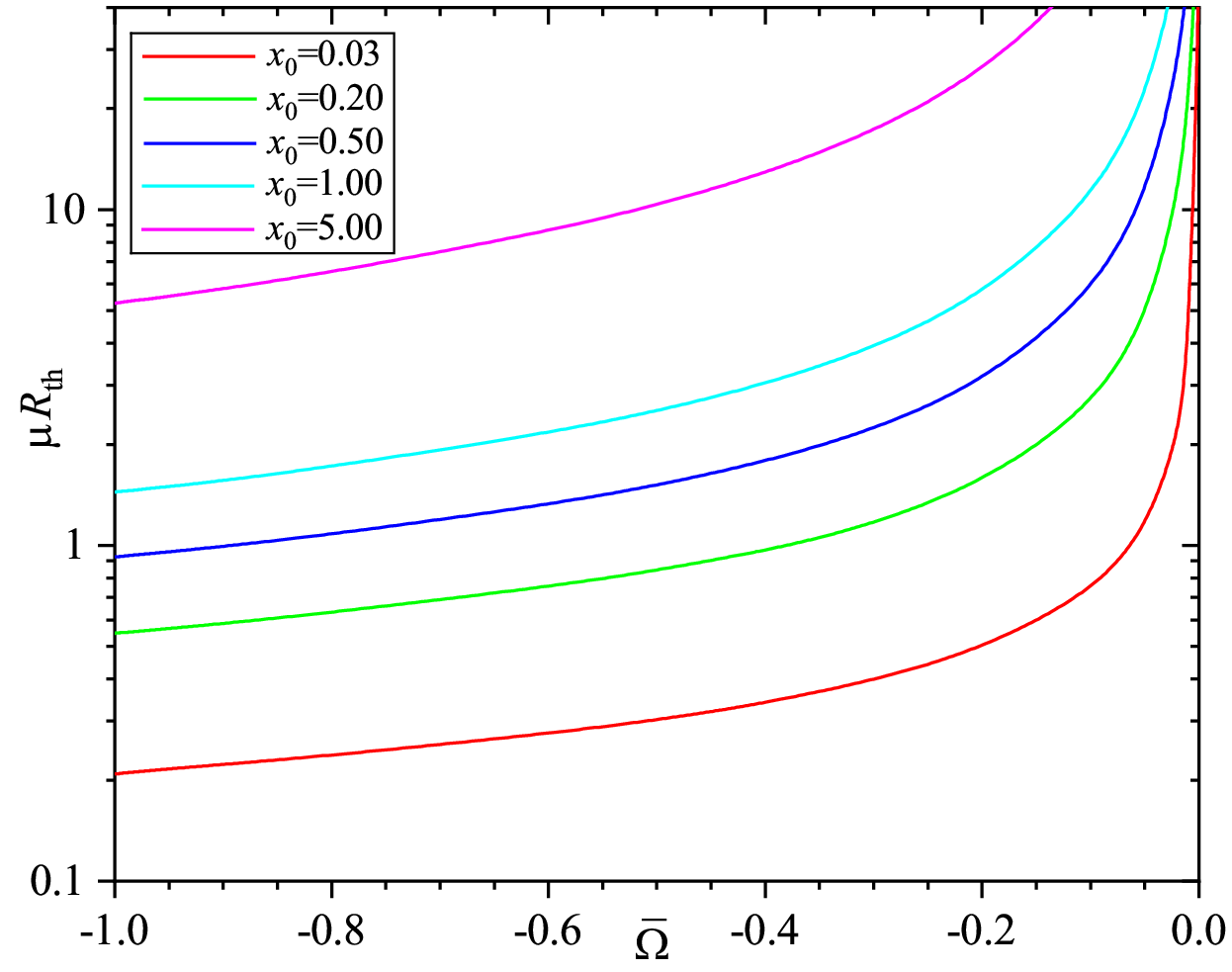}
    \end{center}
    \vspace{-.5cm}
    \caption{The dimensionless circumferential radius of the throat $\bar R_{\text{th}}\equiv \mu  R_{\text{th}}$
as a function of the parameter $\bar\Omega$.
       }
    \label{fig_Rth_Omega}
\end{figure}

To begin with, let us consider the case of an uncharged Dirac field when $\bar e=0$. 
Fig.~\ref{fig_Mass_Omega_e_0} shows the dependencies of the masses $\bar{M}_+$ and $\bar{M}_-$ [calculated using Eqs.~\eqref{expres_mass} or \eqref{m_current_dmls}]
on the frequency $\bar{\Omega}$ for different values of the throat parameter $x_0$.
For all the values of $x_0$ under consideration, the mass curves have a qualitatively similar behavior: 
the configurations have masses approximately equal to 0 as $\bar{\Omega}\to -1$, and, depending on the value of~$x_0$,
the masses $\bar{M}_\pm$ can even be negative (see the insets in Fig.~\ref{fig_Mass_Omega_e_0}); this is caused by the negative energy density of the 
spinor field defined by the expression~\eqref{Ttt_dmls}. 

As  $\bar{\Omega}$ increases, the masses grow smoothly, and in the limit $\bar{\Omega}\to -0$ they demonstrate a fast increase  according to the law $\bar{M}_{\pm}\sim |\bar{\Omega}|^{-1}$.
In this limit: (i)~the masses  $\bar{M}_+$ and $\bar{M}_-$ become equal; (ii)~the metric function $g_{tt}\equiv e^A \to 0$, the metric function $B\approx 1$,
and the dimensionless Kretschmann scalar $\bar{K}\equiv K/\mu^4=\bar R_{\alpha\beta\mu\nu}\bar R^{\alpha\beta\mu\nu}$ is practically equal to 0 throughout all of space;
and (iii)~for all the solutions, the ratio  $\bar{Q}/\bar{M}_\pm \to 1$ from above.

All the systems considered have one throat with the circumferential radius $\bar R_{\text{th}}$, 
and the throat is always located to the left of the center $x=0$ (because of the asymmetry of the solutions). The dependence of $\bar R_{\text{th}}$ on the spinor frequency
 $\bar{\Omega}$ for different values  of the throat parameter $x_0$ is given in Fig.~\ref{fig_Rth_Omega}. By comparing these curves with the corresponding graphs for the masses  $\bar{M}_\pm$
 shown in~Fig.~\ref{fig_Mass_Omega_e_0}, it is seen that for small values of $|\bar{\Omega}|$ the magnitudes of the masses and sizes of the throat are almost the same.
Accordingly, taking into account the expression for the mass~\eqref{m_current_dmls}, it turns out that the contributions to the total mass coming from the mass associated with the throat
 [the first term in Eq.~\eqref{m_current_dmls}] and the mass ensured by the matter (electric and spinor fields) are equal. On the other hand, when $\bar{\Omega}\to -1$, 
 the situation changes drastically: here the masses $\bar{M}_\pm$ tend to 0 (and even become negative), but the throat radius
  $\bar R_{\text{th}}$ always remains essentially nonzero.
  Taking into account the expression for the mass \eqref{m_current_dmls}, from this it follows that in this case the integral yields  a negative contribution which is ensured by the negative energy density
  of the spinor field (the violation of the weak energy condition).

\subsubsection{The case of $\bar{e}\neq0$}
\label{sec_e_neq_0}

\begin{figure}[t]
    \begin{center}
        \includegraphics[width=.49\linewidth]{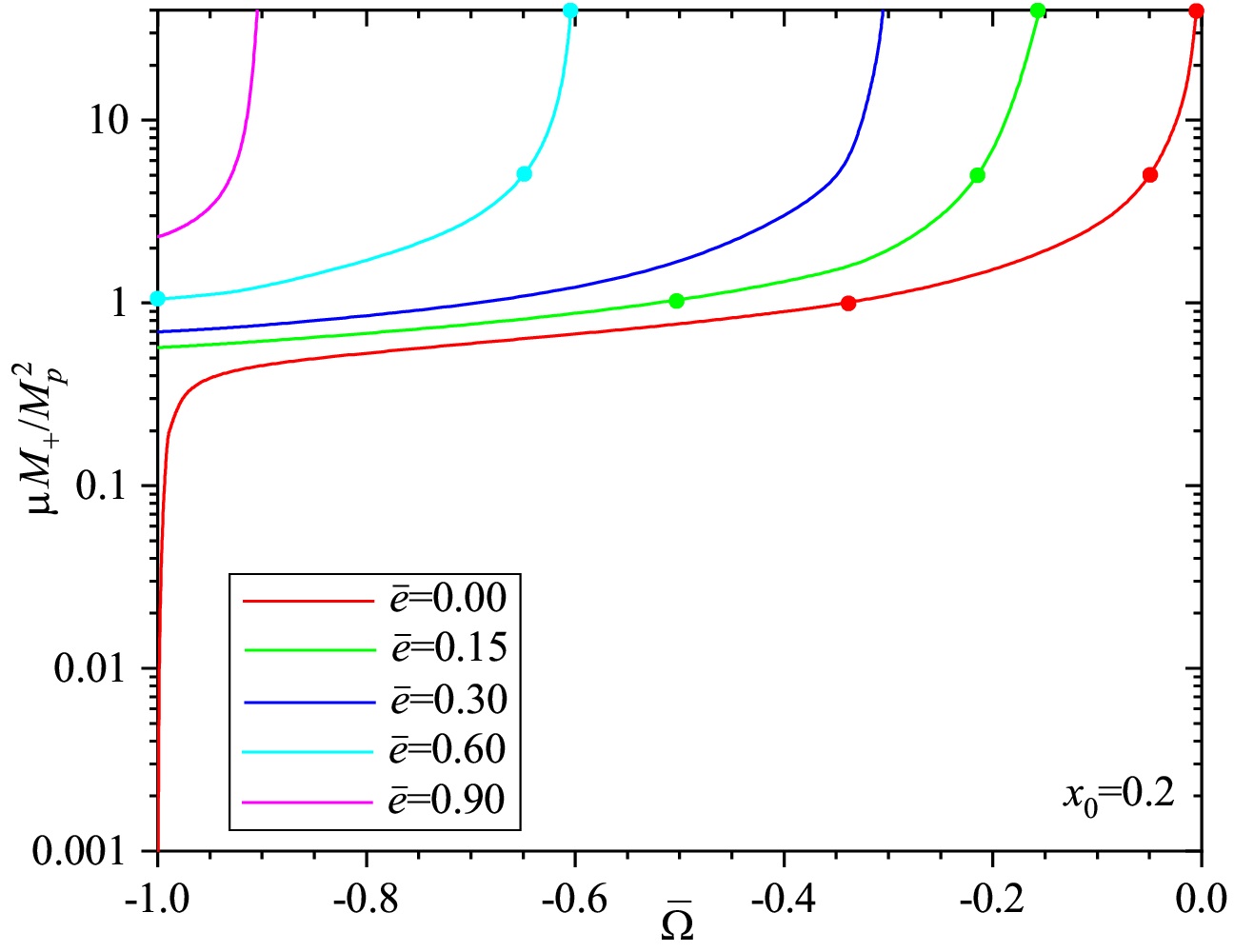}
        \includegraphics[width=.49\linewidth]{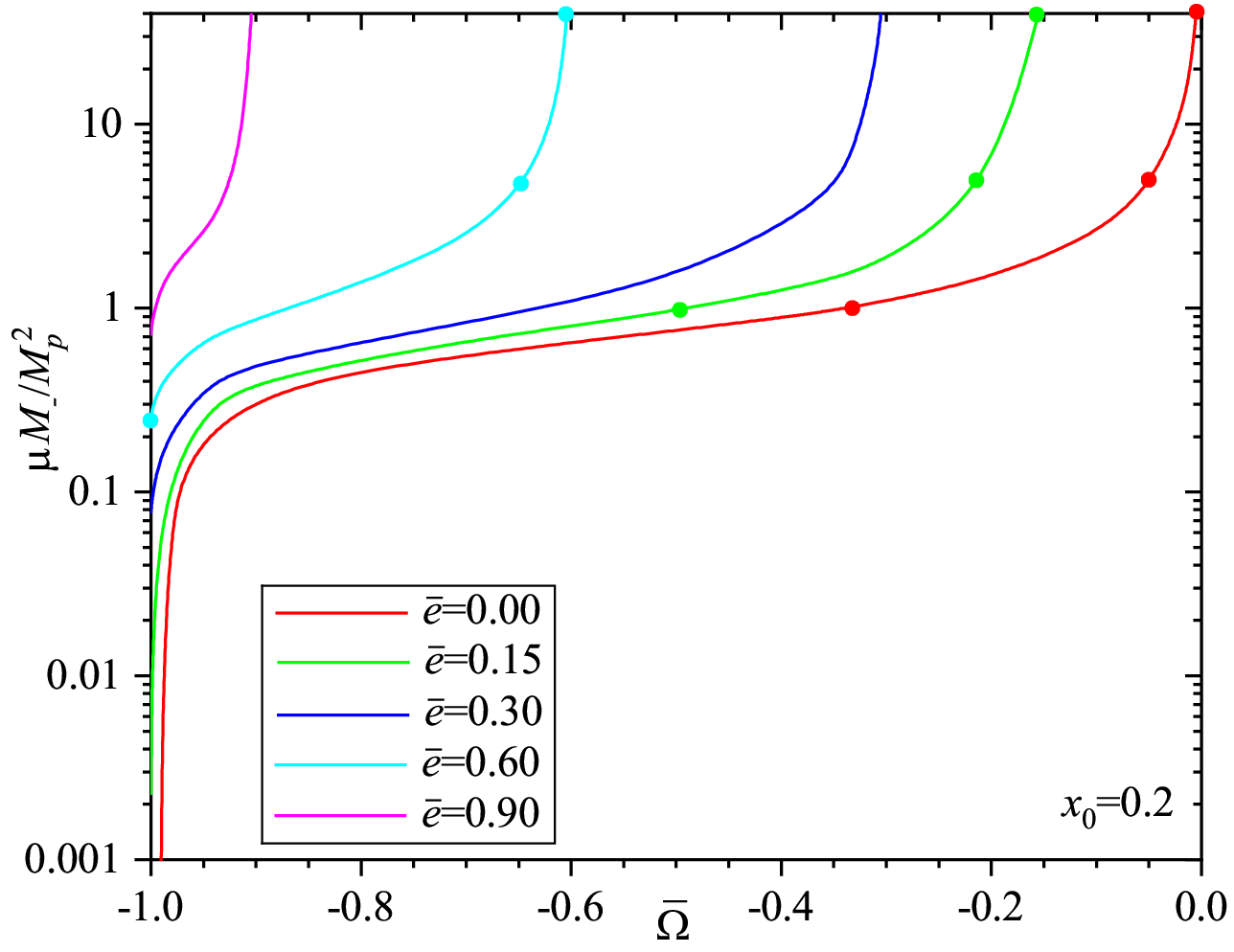}
    \end{center}
    \vspace{-.5cm}
    \caption{The dimensionless total masses $\bar M_\pm$ as functions of
the parameter $\bar\Omega$ for a charged spinor field with different values of the coupling constant~$\bar{e}$. 
As   $\bar\Omega\to \bar{\Omega}_{\text{crit}}\approx -\bar e$, the masses  diverge, as in the case with $\bar{e}=0$ shown in Fig.~\ref{fig_Mass_Omega_e_0}.
The bold dots correspond to the configurations for which the solutions are displayed in Fig.~\ref{fig_plots_sols}.}
    \label{fig_Mass_Omega_e_neq_0}
\end{figure}

\begin{figure}[!]
    \begin{center}
        \includegraphics[width=.5\linewidth]{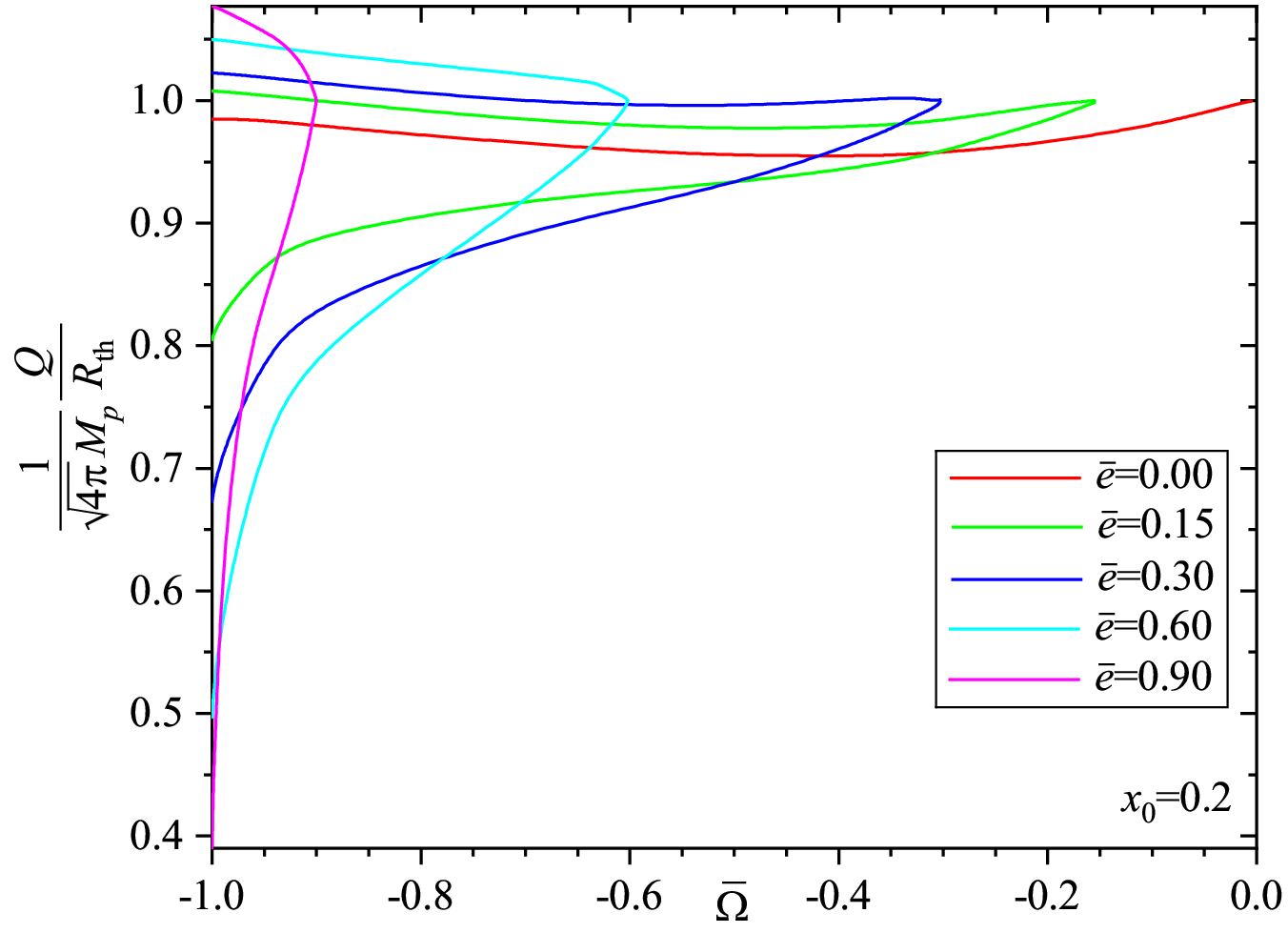}
    \end{center}
    \vspace{-.5cm}
    \caption{The ratio of the charges of the configurations to the radius of their throat $\bar{Q}_\pm/\bar{R}_{\text{th}}$ as a function of  $\bar\Omega$ for different values of $\bar{e}$. 
   For the systems with $\bar{e}= 0$, the charges $\bar{Q}_+$ and $\bar{Q}_-$ are equal, $\bar{Q}_+=\bar{Q}_-= \bar{Q}$.
   For the systems with $\bar{e}\neq 0$, there are turning points located at $\bar{\Omega}=\bar{\Omega}_{\text{crit}}$ where $\bar{Q}_\pm/\bar{R}_{\text{th}}\to 1$:
    the upper parts of the curves correspond to the charge $\bar{Q}_+$, while the lower parts are for the charge $\bar{Q}_-$.
       }
    \label{fig_Q_Rth}
\end{figure}

\begin{figure}[!]
    \begin{center}
        \includegraphics[width=1.\linewidth]{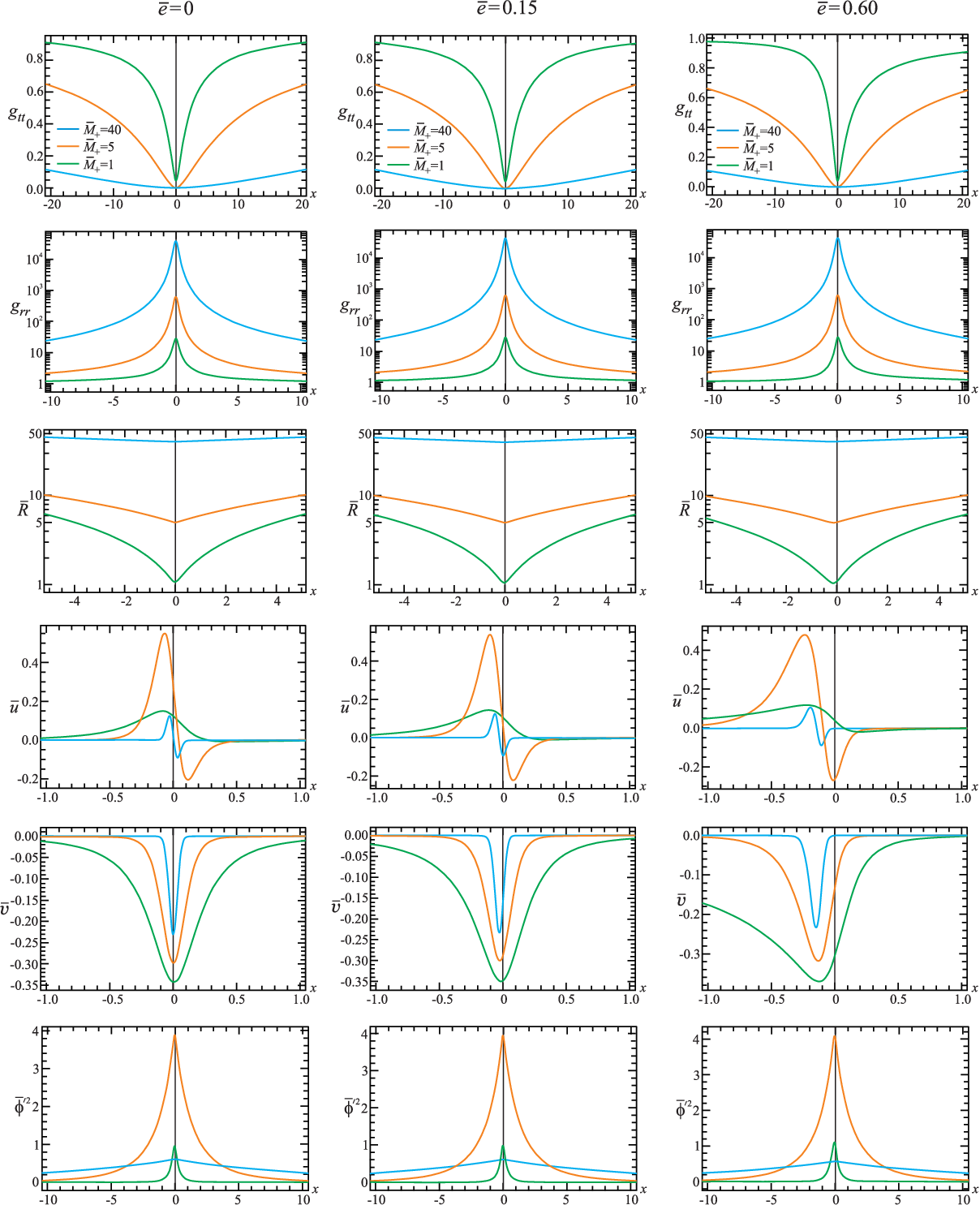}
    \end{center}
    \vspace{-.5cm}
    \caption{Typical solutions for different values of the coupling constant $\bar{e}$ for a fixed $x_0=0.2$.
    The solutions are given for three masses $\bar{M}_+\approx 1, 5, 40$ (the corresponding points in the mass curves for $\bar{M}_\pm$ are shown by bold dots in Fig.~\ref{fig_Mass_Omega_e_neq_0}).
    To visualize the results, we have introduced the rescalings
     $\bar{u}\to \alpha \bar{u}, \bar{v}\to \beta \bar{v}$, and $\bar{\phi}^{\prime 2}\to \gamma \bar{\phi}^{\prime 2}$, where $\alpha=1, 50, 500$, $\beta=1, 5, 50$, $\gamma=1, 100, 1000$
    for $\bar{M}_+\approx 1, 5, 40$, respectively.
       }
    \label{fig_plots_sols}
\end{figure}

Let us now turn to a consideration of systems with nonzero coupling constant. 
To demonstrate a characteristic behavior of the solutions and changes in physical parameters of such systems depending on
the value of $\bar e$, here we restrict ourselves to a consideration of only one fixed value of the throat parameter $x_0=0.2$. 
In this case, varying $\bar\Omega=\bar{U}_{-\infty}$ for each fixed value of $\bar e$,
we obtain families of configurations harboring a wormhole at their core. In doing so, we verify that
 the constraint equation~\eqref{eq_constr} is satisfied (this is achieved by an appropriate choice of the asymptotic value $\bar{U}_{+\infty}$).

Fig.~\ref{fig_Mass_Omega_e_neq_0}  shows the dependencies of the  total masses 
$\bar M_\pm\equiv \mu M_\pm/M_p^2$ [calculated using Eqs.~\eqref{expres_mass} or \eqref{m_current_dmls}] on $\bar \Omega$  for different
values of the coupling constant $\bar e$. It is seen from this figure that as $\bar e$ increases, the mass curves shift to the left of the mass curve for the uncoupled case.
In this case when  $\bar{\Omega}\to -1$ the configurations with $\bar{e}\neq 0$, in contrast to the systems with $\bar{e}=0$, 
may already have the masses $\bar{M}_\pm$ which differ considerably from 0;
also, the masses gradually increase with increasing $\bar\Omega$ and the magnitude of the coupling constant. 
As a result, for every value of $\bar e$, there is some critical value  $\bar{\Omega}_{\text{crit}}\approx -\bar e$ for which, as in the case with $\bar{e}=0$, 
a fast increase in mass occurs (the mass demonstrates a divergent behavior). In this case the behavior of other characteristics of the systems under consideration
remains similar to that of the case with $\bar{e}=0$ (see the items (i)-(iii) from Sec.~\ref{sec_e_0}). Notice also that it is evident from the behavior of the 
mass curves that when $\bar e \to 1$ the corresponding mass curve will degenerate. Thus the type of solutions considered in the present paper is only possible in the range  $0\leq \bar{e}<1$.

It is seen from Fig.~\ref{fig_Q_Rth} that when the coupling constant $\bar{e}$ is zero, the ratio of the charge of the system to the radius of the throat depends only slightly on the spinor frequency
 $\bar{\Omega}$ and is always of the order of  1. However, when $\bar{e}\neq 0$ the situation changes drastically: as  the coupling constant increases, the ratio
  $\bar{Q}_+/\bar{R}_{\text{th}}$ becomes  increasingly large as $\bar{\Omega}$ decreases, while the ratio  $\bar{Q}_-/\bar{R}_{\text{th}}$ becomes smaller and smaller. 
  On the other hand, as  $\bar \Omega \to \bar \Omega_{\text{crit}} \approx -\bar{e}$, we have $\bar{Q}_\pm/\bar{R}_{\text{th}}\to 1$;
this corresponds to the extremal Reissner-Nordstr\"{o}m solution with $\bar{Q}_\pm/\bar{M}_\pm \to 1$ (see Sec.~\ref{sec_approx_sol}).

The results of numerical calculations are exemplified by Fig.~\ref{fig_plots_sols} where typical spatial distributions of the field functions for the systems with three different $\bar e$ are shown.
This figure demonstrates characteristic changes in the distribution  of the spinor fields when, as  the coupling constant $\bar{e}$ increases, their maxima/minima shift to the left.
This results in the fact that the location of the throat (minimum of the metric function $\bar{R}$) shifts to the left as well, and such systems have essentially different values of the masses
 $\bar{M}_+$ and $\bar{M}_-$ (see Fig.~\ref{fig_Mass_Omega_e_neq_0}). Notice also that although for any value of $\bar{e}$, as the total mass of the systems under consideration increases,
 in the vicinity of the throat, the metric function $g_{tt}\equiv e^A \to 0$ and the metric function $g_{rr}\equiv B e^{-A}$ demonstrates a fast growth, the Kretschmann scalar $\bar{K}\to 0$ throughout all of space.

\subsubsection{Energy conditions}

The violation of the null  and weak energy conditions, needed for ensuring nontrivial topology in the system, 
 implies the violation of the following inequalities,
$$
    T_{\mu\nu} k^\mu k^\nu \geq 0 \quad \text{and}\quad T_{\mu\nu} V^\mu V^\nu \geq 0 ,
$$
for any null vector $k^\mu$, $g_{\mu\nu}k^\mu k^\nu =0$, 
and for any timelike vector $V^\mu$, $g_{\mu\nu}V^\mu V^\nu >0$, 
respectively (for a review, see, e.g., Ref.~\cite{Visser}).
The weak energy condition also implies $T_0^0\geq 0$.

Since the violation of the null energy condition implies the violation of the weak and the strong energy conditions, here
we address only the null energy condition. The latter can be reexpressed by making use of the Einstein equations~\eqref{feqs_10} in the form
$$
G_t^t-G_r^r \geq 0 \quad \text{and} \quad G_t^t-G_\theta^\theta \geq 0 .
$$
The null energy condition is violated when one or both of these conditions do not hold in some region of spacetime.
Note that for the configurations considered above the first of these conditions is always violated,
and the second condition is fulfilled for small $|\bar{\Omega}|$, but as  $|\bar{\Omega}|$ increases, it may be violated as well
(the concrete values of $\bar{\Omega}$ for which the violation of  the null energy condition occurs depend on the magnitudes of the parameters $x_0$ and $\bar{e}$).

\subsubsection{Approximate solutions}
\label{sec_approx_sol}

Since the spinor fields are concentrated mainly in the vicinity of the throat 
and  decrease exponentially with distance according to the asymptotic law~\eqref{u_v_asympt} (see also Fig.~\ref{fig_plots_sols}), 
there is some boundary point $x_b$
where the functions $\bar{u}$ and $\bar{v}$ and their derivatives go to zero (let us refer to such solutions as interior ones).
Also, the numerical calculations indicate that  the metric function $B$ for $x > x_b$ is approximately equal to 1, decreasing as $1/x^2$ according to the asymptotic law~\eqref{A_asympt}. 
In turn, the presence of the long-range electric field leads to the fact that, apart from the interior solutions for the spinor and
electric fields, there is also an exterior solution for the electric field. To obtain it, we employ Eqs.~\eqref{Eq_A} and \eqref{Eq_Maxw}
by putting $\bar{u},\bar{v}\approx 0$ and $B\approx 1$ in them.
This enables us to find approximate analytical solutions to the Eqs.~\eqref{Eq_A} and \eqref{Eq_Maxw} in the form
\begin{equation}
A_\pm\approx A_{1\pm}-2 \ln \left\{\cos\left[\arctan\left(\frac{x}{x_0}\right)+ A_{2\pm}\right]\right\}, \quad
\bar{\phi}_\pm\approx \bar{\phi}_{\pm\infty}-\frac{\bar{Q}_\pm}{x-x_0/\tan A_{2\pm} } .
\label{approx_sol}
\end{equation}
Here $A_{1\pm}$ and $A_{2\pm}$ are integration constants related by the expression $A_{1\pm}=2 \ln \left[\cos\left(\pm \pi/2+A_{2\pm}\right)\right]$
(this relation follows from the condition $A_\pm\to 0$ as $x\to \pm\infty$), and the integration constants
$\bar{\phi}_{+\infty}=\left(\bar{U}_{+\infty}-\bar{U}_{-\infty}\right)/\bar{e}$ and $\bar{\phi}_{-\infty}=0$;
the plus/minus signs correspond to the solutions to the right and to the left of the throat, respectively.  

Next, transforming from the metric \eqref{metric} to Schwarzschild-type coordinates, 
$$
ds^2=e^{A(R)}dt^2-B(R)e^{-A(R)} \left(\frac{dr}{dR}\right)^2 dR^2-R^2 \left(d \theta^2 + \sin^2 \theta d \varphi^2\right) ,
$$
and using the expression for the metric function $A$ from Eq.~\eqref{approx_sol}, one can show that, as it should be, the external solution corresponds to 
 the Reissner-Nordstr\"{o}m metric:
$$
g_{tt}\equiv e^{A(R)_\pm}\approx 1\mp\frac{2\bar{M}_\pm}{R}+\frac{\bar{Q}^2_\pm}{R^2} .
$$

Note that for all the configurations considered in Sec. \ref{sec_e_0} and \ref{sec_e_neq_0}, it is typical that, as $\bar \Omega \to \bar \Omega_{\text{crit}} \approx -\bar{e}$,
throughout all of space, the spinor fields $\bar{u}$ and $\bar{v}$ are approximately equal to 0, while the metric function  $B\approx 1$. Accordingly, for such systems, the approximate solutions~\eqref{approx_sol} 
are already valid throughout all of space, including the region in the vicinity of the throat. In this case $\bar{Q}_\pm\to \bar{M}_\pm$; this corresponds to the extremal Reissner-Nordstr\"{o}m solution.

In the case of $\bar{e}\to 0$, when $\bar{U}_{+\infty}\to \bar{U}_{-\infty}$ as well, the first term in the expression for $\bar{\phi}$ from Eq.~\eqref{approx_sol} 
is calculated by evaluating the indeterminate form~$0/0$.

\subsubsection{Stability and the mass-radius relation}
\label{sec_M_R_stab}

Let us now briefly address the question of stability of the systems under consideration. 
The study of the time-dependent Einstein-Dirac-Maxwell model performed in Ref.~\cite{Kain:2023ann}
revealed that, in the course of time, a black hole develops in a system of Ref.~\cite{Konoplya:2021hsm}, and correspondingly such wormholes can not be thought of as traversable.
Since in the present paper we study systems of the type~\cite{Konoplya:2021hsm}, one can expect that such configurations will also be unstable,
at least for some values of the system parameters. Nevertheless, to answer this question definitively, it is necessary to perform 
a special analysis of linear and nonlinear perturbations, analogous to what was done for the systems supported by
a ghost scalar field~\cite{Gonzalez:2008wd,Gonzalez:2008xk,Gonzalez:2009hn,Blazquez-Salcedo:2018ipc}.
 However, the presence of an asymmetric spinor field greatly complicates such a stability analysis.

Nevertheless, even without performing an ambitious stability analysis, it is clear that, for stable configurations, the binding energy (BE)
must necessarily be positive (energy stability). 
Consistent with this, let us  calculate the BE,
which is defined as the difference between the energy of $N_f$ free particles, ${\cal E}_f=N_f \mu$, and the total energy of the system, ${\cal E}_t=M$,
that is, 
 \begin{equation}
\label{BE_gen}
\text{BE}={\cal E}_f-{\cal E}_t .
\end{equation}
Here the total particle number $N_f$ can be associated with the Noether charge of the spinor field  $Q_N$, which 
becomes an integer after quantisation, $Q_N=N_f$. The Noether charge
is defined via the timelike component of the four-current $j^\alpha=\bar \psi \gamma^\alpha \psi$
as
$
Q_N=\int\sqrt{-\cal{g}} j^t d^3 x ,
$
where in our case $\sqrt{-\cal{g}}j^t = B^{3/2}e^{-3 A/2}\left(r^2+r_0^2\right) \sin{\theta} \left(\psi^\dag \psi\right)$.
For symmetric systems, the lower limit in the above integral is the point $r=0$ where a throat or an equator are located.

However, for the asymmetric systems under investigation, the situation is more complicated, 
since in such systems the throat is located in general asymmetrically with respect to the center.
In this case the lower limit in the above integral is a point $r=r_{\text{max}}$ where the maximum of the charge density of the spinor field 
 $j^t=\bar{\psi}\gamma^0\psi=e^{-A/2}\left(\psi^\dag \psi\right)$ is located~\cite{Dzhunushaliev:2025fbf}.
In terms of the dimensionless variables \eqref{dmls_var}, we then have for the Noether charge:
\begin{equation}
\label{part_num}
N_{f\pm}=Q_{N\pm}=\pm \left(\frac{M_p}{\mu}\right)^2\int_{x_{\text{max}}}^{\pm\infty} B^{3/2}e^{-3 A/2}\left(x^2+x_0^2\right)\left(\bar u^2+\bar v^2\right) dx.
\end{equation}

\begin{figure}[t]
    \begin{center}
        \includegraphics[width=.49\linewidth]{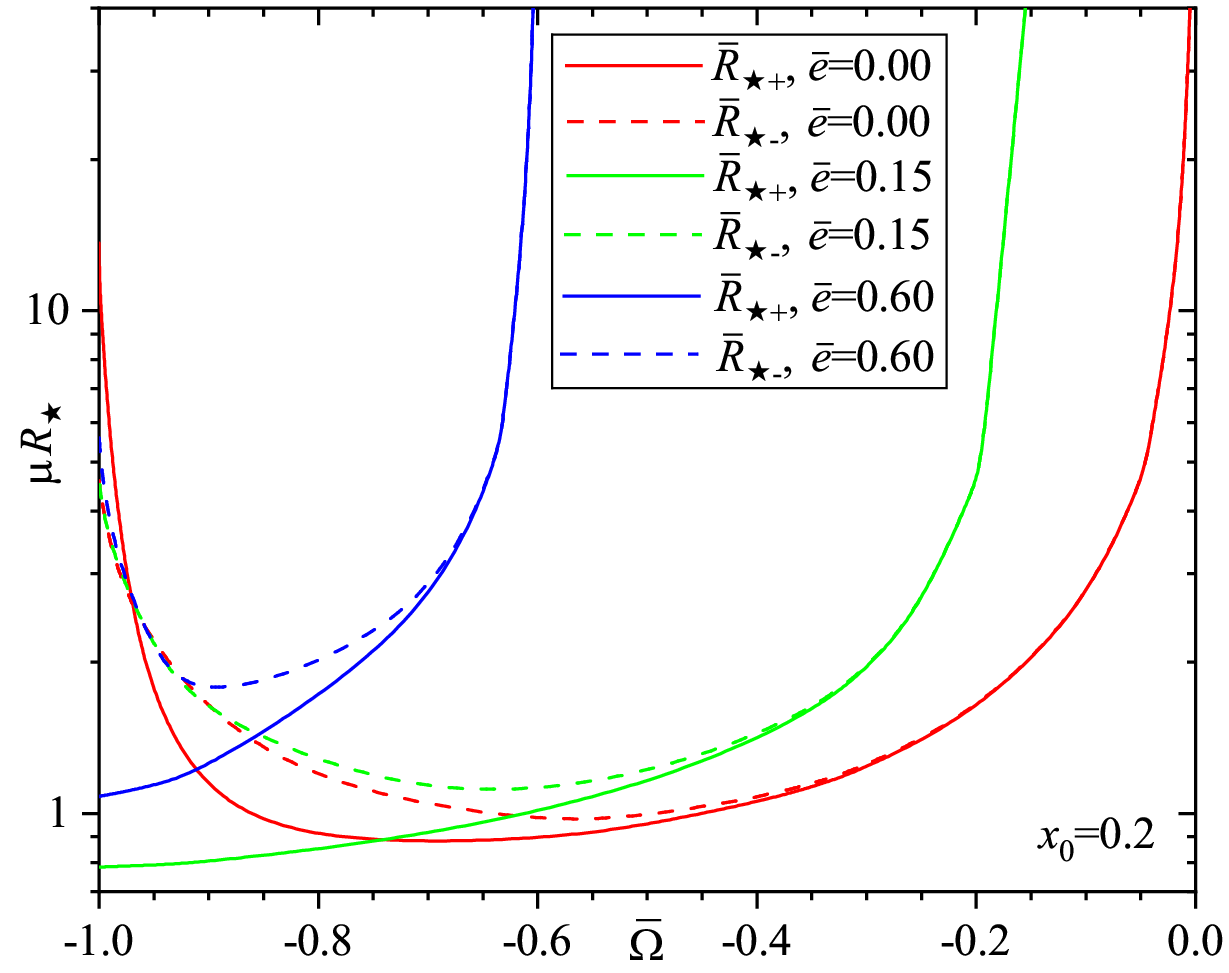}
        \includegraphics[width=.49\linewidth]{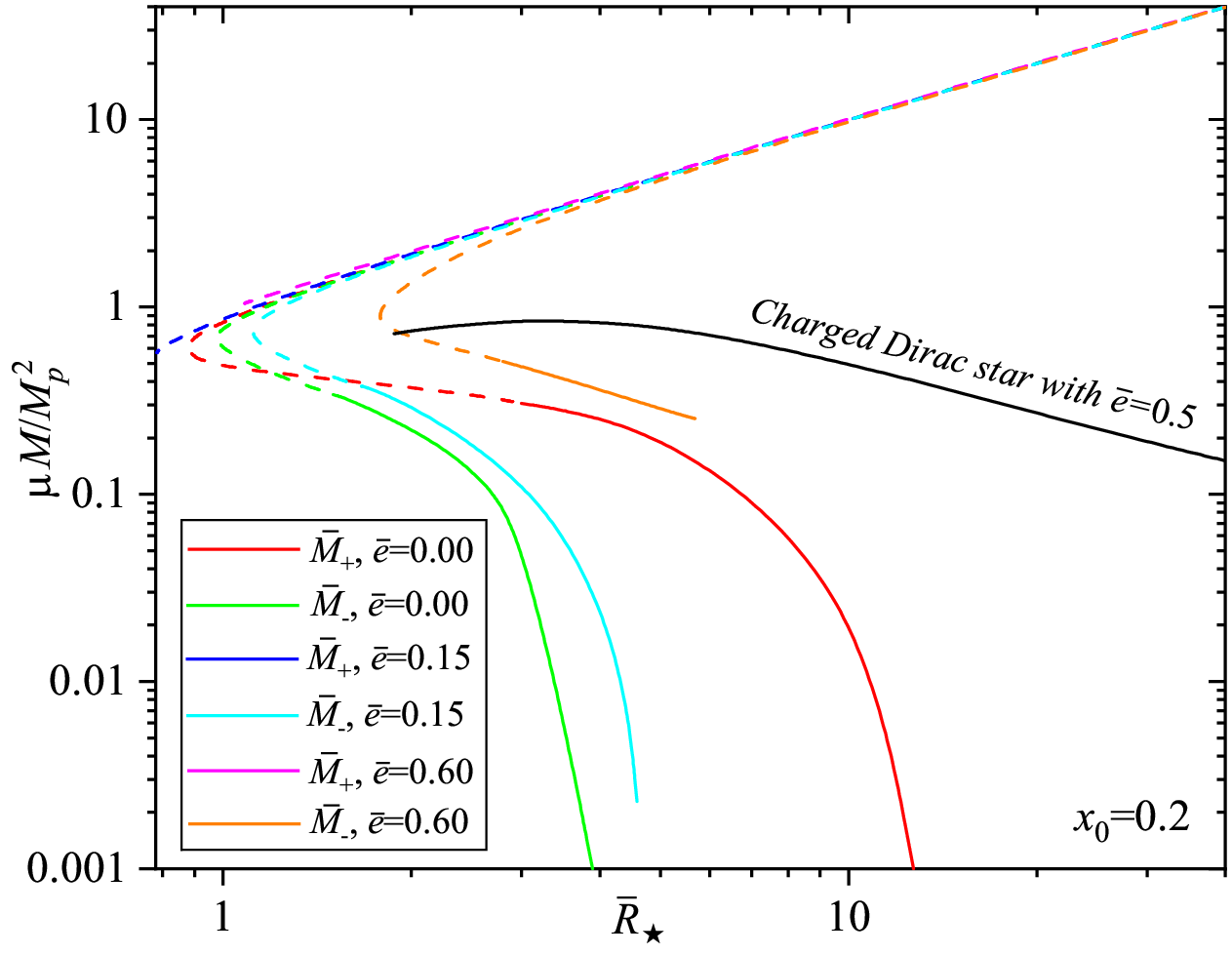}
    \end{center}
    \vspace{-.5cm}
    \caption{Left panel: the dimensionless effective radius $\bar R_{\star}\equiv \mu R_{\star}$ as a function of
the parameter $\bar\Omega$ for the configurations with different values of $\bar{e}$. 
Right panel: the mass-radius relations for the configurations with different values of $\bar{e}$. 
The dashed lines mark the parts of the mass-radius curves where the binding energy is negative (energetically unstable systems).
The solid black line represents the mass-radius curve for Einstein-Dirac-Maxwell systems with trivial spacetime topology.
}
    \label{fig_R_eff_Omega}
\end{figure}

A necessary condition for the energy stability of the systems under consideration is the positivity of the BE. In this connection, in
the mass curves given in Fig.~\ref{fig_R_eff_Omega}, we depict the parts of the curves where the BE is negative by dashed lines:
the corresponding configurations are certainly unstable.

Another known stability criterion involves a consideration of behavior of mass-radius curves; it is used in considering 
strongly gravitating systems, such as, for example, neutron~\cite{HPY} and boson~\cite{Schunck:2003kk,Liebling:2012fv} stars. 
Above we have found the masses of the systems under investigation for different values of the throat parameter $x_0$ and of the coupling constant~$\bar{e}$.
To construct the mass-radius curves, it is now necessary to determine the radius of the configurations under consideration. Since such systems do not possess a sharp surface, 
their sizes are not uniquely defined. Here we adopt the following definition for the effective radius that 
is rather insensitive to the various definitions employed~\cite{Kleihaus:2011sx,Dzhunushaliev:2014bya}:
 \begin{equation}
\label{R_max_def}
R_{\star\pm}=\frac{\int_{x_{\text{max}}}^{\pm\infty}\sqrt{-\cal{g}} j^t R(r) dr}{\int_{x_{\text{max}}}^{\pm\infty}\sqrt{-\cal{g}} j^t dr} ,
\end{equation}
where $R$ is defined by Eq.~\eqref{circ_radius} and the expression for the timelike component of the four-current
$j^t$ is given below Eq.~\eqref{BE_gen}. In turn, the definition of  the lower limit in the integrals can be found above Eq.~\eqref{part_num}.

Using the expression~\eqref{R_max_def}, in the left panel of Fig.~\ref{fig_R_eff_Omega},
we have plotted the dependence of the effective radius  on the spinor frequency $\bar{\Omega}$ for different values of the coupling constant $\bar{e}$. 
This figure shows the strong dependence of the character of qualitative behavior of the radius on whether the spinor field is charged or neutral.
When $\bar\Omega\to \bar{\Omega}_{\text{crit}}\approx -\bar e$, the radii $R_{\star +}$ (the sizes of the systems located to the right of the throat) and $R_{\star -}$ (the sizes of the systems located to the left of the throat)
are always equal to each other; in turn, when $\bar\Omega\to -1$, these radii (i)~differ considerably in magnitude; 
and~(ii)~for the neutral field, the radius $R_{\star +}$ increases rapidly, but for the charged field it decreases smoothly to some nonzero value. 

Next, using the expressions \eqref{expres_mass} or \eqref{m_current_dmls} and \eqref{R_max_def}, we have constructed the mass-radius relations given in the right panel of Fig.~\ref{fig_R_eff_Omega}.
The curves shown here correspond to all configurations for which regular solutions have been found in the range of the spinor frequencies $-1<\bar\Omega<\bar{\Omega}_{\text{crit}}$.
It is seen from these graphs that, depending on the value of the coupling constant $\bar{e}$,  the curves may have (or not have) turning points. In the case of neutron and boson stars,
a turning point usually divides stable and unstable configurations; 
a characteristic feature of stable systems is that their masses increase with decreasing the radii. 
For the configurations with a nontrivial spacetime topology considered here, the situation is more complicated: 
the mass-radius curves either may have no turning points at all (the systems with $\bar{e}=0.15$ and $\bar{e}=0.60$ located to the right of the throat) or,
in the regions where the mass increases with decreasing the radius, the binding energy can be negative (energetically unstable systems).   

In turn, it is of interest to compare the mass-radius relations found here with those obtained for localized configurations with trivial spacetime topology. 
For this purpose, in the right panel of Fig.~\ref{fig_R_eff_Omega}, we have shown the corresponding dependence for a charged Dirac star with trivial spacetime topology obtained
within Einstein-Dirac-Maxwell theory (see Ref.~\cite{Dzhunushaliev:2019kiy}). It is seen from this graph that the behavior of the mass-radius curve 
for the system with trivial topology differs considerably (both qualitatively and quantitatively) from that of the systems with nontrivial topology and close value of the coupling constant $\bar{e}=0.6$
shown by the corresponding curves for $\bar{M}_+$ and  $\bar{M}_-$.

\section{Conclusion}
\label{concl}

In the present paper we have studied gravitating spherically symmetric systems possessing a wormhole topology and consisting of a Maxwellian electric field and
two spinor fields having opposite spins. 
 In contrast to the configurations of Ref.~\cite{Konoplya:2021hsm}, here we have considered a situation where a  wormhole 
 connects two  identical Minkowski spacetimes.
 For such a system, we have constructed complete families of regular, asymptotically
flat solutions for explicitly time-dependent spinor fields (charged and neutral) 
oscillating with a frequency~$\bar\Omega$ whose values lie in the entire allowed range $-1<\bar\Omega<\bar{\Omega}_{\text{crit}}$. 
It is demonstrated that the type of solutions considered in the present paper is only possible in the range of values of the coupling constant $0\leq \bar{e}<1$.

The resulting configurations are asymmetric with respect to the center, including the asymmetry of the 
$g_{tt}$-component of the metric tensor (unlike the systems of Ref.~\cite{Konoplya:2021hsm}). 
Because of the asymmetry,  masses and sizes of the configurations observed at the two asymptotic ends of the wormhole (that is, when $x\to \pm \infty$) 
may differ considerably.
It is shown that in the case with the coupling constant $\bar{e}=0$, depending on the value of the throat parameter~$x_0$, 
these solutions describe configurations that may possess both positive and negative ADM masses. 

Let us enumerate the most interesting features of the configurations under consideration: 
\begin{itemize}
\item[(i)] In the case of an uncharged spinor field ($\bar{e}=0$), whatever the value of the throat parameter $x_0$, 
there is a qualitatively similar behavior of the dependencies of the ADM masses on the spinor frequency
 $\bar\Omega$: as $\bar\Omega \to -1$, the masses are approximately equal to 0, and can even be negative;
 as $\bar\Omega \to -0$, the masses increase according to the law  $\bar{M}_{\pm}\sim \bar{|\Omega|}^{-1}$. 
 In this case, as  $\bar\Omega$ increases from~-1  to 0, 
 the masses increase smoothly (with no turning points), unlike the systems with spinor fields in which a wormhole topology is ensured 
 by a ghost scalar field~\cite{Hao:2023igi,Dzhunushaliev:2025fbf,Dzhunushaliev:2025qhw}.

\item[(ii)] In the case of a charged spinor field ($\bar{e}\neq 0$), the behavior of the mass curves resembles in general the case of the systems with $\bar{e}=0$:
here there is also some critical value   $\bar{\Omega}_{\text{crit}}\approx -\bar e$ for which, as in the case with  $\bar{e}=0$, 
there is a fast increase in mass which demonstrates a divergent behavior as $\bar \Omega \to \bar \Omega_{\text{crit}}$.
However, unlike the systems with $\bar{e}=0$,  as $\bar{\Omega}\to -1$, the configurations with $\bar{e}\neq 0$
may already have a mass which differs considerably from~0 and gradually increases with increasing  $\bar\Omega$ and the magnitude of the coupling constant.

\item[(iii)] It is typical of the configurations from the items (i) and (ii) that, as $\bar \Omega \to \bar \Omega_{\text{crit}} \approx -\bar{e}$, in all of space 
the spinor fields  $\bar{u}$ and $\bar{v}$ are approximately equal to zero, while the metric function $B\approx 1$. 
For such systems, it is possible to find approximate analytical solutions in the form~\eqref{approx_sol}, which
are already valid throughout all of space, including the region near the throat. 
These solutions correspond  to the Reissner-Nordstr\"{o}m solution, and when the charges of the system $\bar{Q}_\pm\to \bar{M}_\pm$, we deal with the extremal Reissner-Nordstr\"{o}m solution.

\item[(iv)] It follows from a consideration of the binding energy and of the behavior of the mass-radius curves  that there can exist energetically stable configurations for all values of the coupling constant $\bar{e}$. 

\item[(v)] Unlike the systems with nontrivial spacetime topology supported by a ghost scalar field and spinor fields~\cite{Dzhunushaliev:2025fbf,Dzhunushaliev:2025qhw},
all the configurations considered in the present paper  have only one throat.  
Nevertheless, since we have considered only a restricted range of values of the throat parameter $x_0$, 
we cannot rule out the possibility that configurations with two or more throats may exist for other values of the input parameters of the system. However, this issue requires special studies.
\end{itemize}

\section*{Acknowledgements}

We gratefully acknowledge support provided by the program No.~AP23490322  (Exploration of Thermodynamic Properties of Relativistic Compact Objects within the Framework of Geometrothermodynamics (GTD))
of the Committee of Science of the Ministry of Science and Higher Education of the Republic of Kazakhstan.


\begin{thebibliography}{99}
  
\bibitem{Visser}
M. Visser,  {\it Lorentzian Wormholes: From Einstein to Hawking} (Woodbury, New York, 1996).  

\bibitem{AmenTsu2010}
 L. Amendola and S. Tsujikawa, {\it  Dark Energy: Theory and Observations} (Cambridge University Press,
Cambridge, England, 2010).

\bibitem{Ade:2015xua}
  P.~A.~R.~Ade {\it et al.} [Planck Collaboration],
  Astron.\ Astrophys.\  {\bf 594}, A13 (2016).

\bibitem{Bronnikov:1973fh}
  K.~A.~Bronnikov,
  Acta Phys.\ Polon.\  {\bf B4}, 251 (1973).

\bibitem{Ellis:1973yv}
  H.~G.~Ellis,
  J.\ Math.\ Phys.\  {\bf 14}, 104 (1973).

\bibitem{Ellis:1979bh}
  H.~G.~Ellis,
  Gen.\ Rel.\ Grav.\  {\bf 10}, 105 (1979).

\bibitem{Kodama:1978dw}
  T.~Kodama,
  Phys.\ Rev.\ D {\bf 18}, 3529 (1978).

\bibitem{Kodama:1978zg}
  T.~Kodama, L.~C.~S.~de Oliveira, and F.~C.~Santos,
  Phys.\ Rev.\ D {\bf 19}, 3576 (1979).

\bibitem{Blazquez-Salcedo:2020czn}
J.~L.~Bl\'azquez-Salcedo, C.~Knoll, and E.~Radu,
Phys. Rev. Lett. \textbf{126},  101102 (2021).

\bibitem{Bolokhov:2021fil}
S.~Bolokhov, K.~Bronnikov, S.~Krasnikov, and M.~Skvortsova,
Grav. Cosmol. \textbf{27},  401 (2021).

\bibitem{Danielson:2021aor}
D.~L.~Danielson, G.~Satishchandran, R.~M.~Wald, and R.~J.~Weinbaum,
Phys. Rev. D \textbf{104},  124055 (2021).

\bibitem{Konoplya:2021hsm}
R.~A.~Konoplya and A.~Zhidenko,
Phys. Rev. Lett. \textbf{128},  091104 (2022).

\bibitem{Wang:2022aze}
Y.~Q.~Wang, S.~W.~Wei, and Y.~X.~Liu,
``Comment on ''Traversable Wormholes in General Relativity'',''
[arXiv:2206.12250 [gr-qc]].

  
\bibitem{Lawrie2002}
I.~Lawrie, {\it A Unified Grand Tour of Theoretical Physics} (Institute of Physics Publishing, Bristol, 2002).

 
 \bibitem{Soler:1970xp}
 M.~Soler,
Phys.\ Rev.\ D {\bf 1}, 2766 (1970).

\bibitem{Li:1982gf}
X.~z.~Li, K.~l.~Wang, and J.~z.~Zhang,
Nuovo Cimento\ A {\bf 75}, 87 (1983).

\bibitem{Li:1985gf}
K.~L.~Wang and J.~Z.~Zhang,
Nuovo Cimento\ A {\bf 86}, 32 (1985).

\bibitem{Herdeiro:2017fhv}
C.~A.~R.~Herdeiro, A.~M.~Pombo, and E.~Radu,
Phys.\ Lett.\ B {\bf 773}, 654 (2017).

\bibitem{Dzhunushaliev:2023sdq}
V.~Dzhunushaliev, V.~Folomeev, and D.~Zholdakhmet,
Eur. Phys. J. C \textbf{83},  550 (2023).

\bibitem{pardiso}N.I.M.~Gould, J.A.~Scott, Y.~Hu,
ACM Trans. Math. Softw. {\bf 33}, 10 (2007);\\
O.~Schenk, K.~Gartner, Future Gener. Comput. Syst. {\bf 20}, 475 (2004).
  
\bibitem{Herdeiro:2019mbz}
C.~Herdeiro, I.~Perapechka, E.~Radu, and Y.~Shnir,
Phys. Lett. B \textbf{797}, 134845 (2019).

\bibitem{Herdeiro:2021jgc}
C.~Herdeiro, I.~Perapechka, E.~Radu, and Y.~Shnir,
Phys. Lett. B \textbf{824}, 136811 (2022).

\bibitem{Charalampidis:2013ixa}
E.~Charalampidis, T.~Ioannidou, B.~Kleihaus, and J.~Kunz,
Phys. Rev. D \textbf{87},  084069 (2013).

\bibitem{Hauser:2013jea}
O.~Hauser, R.~Ibadov, B.~Kleihaus, and J.~Kunz,
Phys. Rev. D \textbf{89},  064010 (2014).

\bibitem{Dzhunushaliev:2014mza}
V.~Dzhunushaliev, V.~Folomeev, B.~Kleihaus, and J.~Kunz,
Phys. Rev. D \textbf{89},  084018 (2014).

\bibitem{Dzhunushaliev:2025fbf}
V.~Dzhunushaliev, V.~Folomeev, and S.~Sakhiyev,
Gen. Rel. Grav. \textbf{57},  126 (2025).

\bibitem{Dzhunushaliev:2025qhw}
V.~Dzhunushaliev, V.~Folomeev, and N.~Saduyev,
J. Phys. Conf. Ser. \textbf{3089},  012007 (2025).

\bibitem{Misner:1964je}
  C.~W.~Misner and D.~H.~Sharp,
  Phys.\ Rev.\  {\bf 136}, B571 (1964).

\bibitem{Kain:2023ann}
B.~Kain,
Phys. Rev. D \textbf{108},  044019 (2023).

\bibitem{Gonzalez:2008wd}
  J.~A.~Gonzalez, F.~S.~Guzman, and O.~Sarbach,
  Class.\ Quant.\ Grav.\  {\bf 26}, 015010 (2009).

\bibitem{Gonzalez:2008xk}
  J.~A.~Gonzalez, F.~S.~Guzman, and O.~Sarbach,
 Class.\ Quant.\ Grav.\  {\bf 26}, 015011 (2009).

\bibitem{Gonzalez:2009hn}
J.~A.~Gonzalez, F.~S.~Guzman, and O.~Sarbach,
Phys. Rev. D \textbf{80}, 024023 (2009).

\bibitem{Blazquez-Salcedo:2018ipc}
J.~L.~Bl\'azquez-Salcedo, X.~Y.~Chew, and J.~Kunz,
Phys. Rev. D \textbf{98}, 044035 (2018).

\bibitem{HPY}
P.~Haensel,  A.~Y.~Potekhin, and D.~G.~Yakovlev, {\it Neutron Stars 1: Equation of State and Structure} (Springer, New York, 2007).

\bibitem{Schunck:2003kk}
  F.~E.~Schunck and E.~W.~Mielke,
  Class.\ Quant.\ Grav.\  {\bf 20}, R301 (2003).

\bibitem{Liebling:2012fv}
  S.~L.~Liebling and C.~Palenzuela,
  Living Rev.\ Relativity\  {\bf 15}, 6 (2012);
   {\bf 20}, 5 (2017).

\bibitem{Kleihaus:2011sx}
B.~Kleihaus, J.~Kunz, and S.~Schneider,
Phys. Rev. D \textbf{85}, 024045 (2012).

\bibitem{Dzhunushaliev:2014bya}
V.~Dzhunushaliev, V.~Folomeev, C.~Hoffmann, B.~Kleihaus, and J.~Kunz,
Phys. Rev. D \textbf{90},  124038 (2014).

\bibitem{Dzhunushaliev:2019kiy}
V.~Dzhunushaliev and V.~Folomeev,
Phys. Rev. D \textbf{99}, 104066 (2019).

\bibitem{Hao:2023igi}
C.~H.~Hao, S.~X.~Sun, L.~X.~Huang, R.~Zhang, X.~Su, and Y.~Q.~Wang,
JCAP \textbf{04}, 057 (2024).

\end{thebibliography}
\end{document}